\documentclass[twocolumn,tighten]{aastex63}

\usepackage{graphicx}
\usepackage{amsmath}
\usepackage{amssymb}
\usepackage{color}
\usepackage{mathtools}
\usepackage{ulem}
\usepackage[all]{hypcap}

\newcommand\msun{\, M_\odot}

\newcommand\be{\begin{equation}}
\newcommand\ee{\end{equation}}

\def\apjl{ApJL}%
\def\apjs{ApJS}%
\def\aap{A\& A}%
\begin{document}

\title{Implication of spin constraints by the Event Horizon Telescope on stellar orbits in the Galactic Center}

\correspondingauthor{Giacomo Fragione}
\email{giacomo.fragione@northwestern.edu}

\author[0000-0002-7330-027X]{Giacomo Fragione}
\affil{Center for Interdisciplinary Exploration \& Research in Astrophysics (CIERA) and Department of Physics \& Astronomy, Northwestern University, Evanston, IL 60208, USA}
\affil{Department of Physics \& Astronomy, Northwestern University, Evanston, IL 60202, USA}

\author[0000-0003-4330-287X]{Abraham Loeb}
\affil{Astronomy Department, Harvard University, 60 Garden St., Cambridge, MA 02138, USA}

\begin{abstract}
The center of the Milky Way hosts the closest supermassive black hole, SgrA$^*$. Decades of near-infrared observations of our Galactic Center have shown the presence of a small population of stars (the so called S-star cluster) orbiting SgrA$^*$, which were recently reported to be arranged in two orthogonal disks. In this case, the timescale for Lense-Thirring precession of S-stars should be longer than their age, implying a low spin for SgrA$^*$. In contrast, the recent results by the Event Horizon Telescope favor a highly-spinning SgrA$^*$, which seem to suggest that the S-stars could not be arranged in disks. Alternatively, the spin of SgrA$^*$ must be small, suggesting that the models for its observed image are incomplete.
\end{abstract}

\section{Introduction}
\label{sect:intro}

Supermassive black holes (SMBHs) are ubiquitous at the center of nearly every galaxy \citep{korm2013}. SgrA$^*$ at the center of our Milky Way is the closest example, representing a unique laboratory to study stellar dynamics and to test general relativity under extreme conditions \citep[e.g.,][]{alexander2017}.

SgrA$^*$ is surrounded by a kaleidoscopic environment, which comprises of young and old stars, compact remnants, gas, and molecular clouds \citep[e.g.,][]{genz2010}. Within $\sim 1$\,pc from the center of the Galaxy, the dynamics of this multitude of astrophysical objects is dictated to leading order by the gravitational potential of the SMBH \citep[e.g.,][]{merritt2013,alexander2017}. Decades of of near-infrared observations of our Galactic Center have shown that a population of about 40 stars (S2 has the shortest orbital period of about $15$ yr), the so called S-star cluster, orbits SgrA$^*$ close enough that it can be used as a dynamical probe of its existence \citep[e.g.,][]{GhezDuchene2003,sch2002Natur,GhezSalim2008,gill2009,gill2017}. These observations have constrained  the mass of our SMBH to about $4\times10^6\msun$, and have tested General Relativity \citep[e.g.,][]{DoHees2019,GravityCollaborationAbuter2020}. However, the spin of SgrA$^*$ remains poorly constrained. 

The Event Horizon Telescope (EHT) has made it possible to study SMBHs with direct imaging \citep{EventHorizonTelescopeCollaborationAkiyama2019}. The mass and spin of SMBHs can be constrained by modeling interferometric EHT data sets with snapshot images of numerical simulations or semianalytic models \citep[e.g.,][]{AgolKrolik2000,BroderickLoeb2005,BroderickLoeb2006,BroderickLoeb2009,dexter2010}. The first direct image of the SMBH at the center of M87 has shown the power of this unprecedented tool \citep{EventHorizonTelescopeCollaborationAkiyama2019b}.

Recently, the EHT collaboration has released the first image of SgrA$^*$, which showed a compact emission region with intrahour variability \citep{AkiyamaAlberdi2022}. Using a large suite of numerical simulations, the image of SgrA$^*$ has been shown to be consistent with the expected appearance of a Kerr black hole with mass of about $4\times10^6\msun$ \citep{AkiyamaAlberdi2022b}, in agreement with the current constraints from individual S-star orbits \citep[e.g.,][]{GhezSalim2008,gill2009}. Moreover, the EHT models disfavor scenarios where the SMBH is viewed at high inclination, with a preference for inclination of about $30^\circ$, as well as disfavor a nonspinning black hole, with a preference for a spin $\chi_{\rm BH} >0.5$ \citep[see Figure 4 in][]{AkiyamaAlberdi2022}.

The preference for rapidly-rotating configurations has important implications for the stellar orbits of the stars around it. By analysing the kinematics of the S-stars, \citet{ali2020} have recently argued that they are arranged in two almost orthogonal disks. \citet{FragioneLoeb2020} have shown that the spin of SgrA$^*$ is then constrained to be $\chi_{\rm BH}\lesssim 0.1$, by requiring that the frame-dragging precession has not had enough time to drive the S-stars out from their disky configuration.

In this Letter, we reanalyze the geometrical argument of \citet{FragioneLoeb2020} in light of the recent results from the EHT, and we discuss the implications of a possibly highly-spinning SgrA$^*$ for the distribution of the stellar orbits around it.

\section{The closest stars to SgrA*}
\label{sect:sstars}

\citet{ali2020} have recently argued that the S-stars are arranged in two orthogonal disks (the so-called ``red'' and ``black'' disks), which are located at a position angle of approximately $\pm 45^\circ$ with respect to the Galactic plane. \citet{peiss2020a,peiss2020b} have also argued for the discovery of six new S-stars, fainter and less massive than the classical S-stars, some of which (in particular S62 on a $9.9$~yr orbit) are even closer to SgrA$^*$ than S2. While the classical S-stars have masses in the range $8\msun$--$14\msun$ \citep{habibi2017}, \citet{peiss2020b} estimated a mass of about $6.1\msun$ for S62 and a mass in the range $2\msun$--$3\msun$ for the other five new candidates.

The advent of the near-infrared GRAVITY instrument at the VLTI has marked the beginning of a new era in observations of the Galactic Center \citep{grav2018b,grav2018a}. GRAVITY can improve the localization of the innermost stars in our Galaxy by a factor of about $20$ compared to adaptive optics and, most importantly, the high angular resolution of GRAVITY allows it to overcome the confusion limit of adaptive optics imaging. In a recent analysis of deep images of the Galactic Center, however, none of the GRAVITY sources matches the $9.9$~yr-orbital period star as reported by \citet{peiss2020a}. Moreover, using GRAVITY data, \citet{vonFellenbergGillessen2022} could not confirm that S-stars in the central region are organized in two orthogonal disks.

\begin{figure} 
\centering
\includegraphics[scale=0.55]{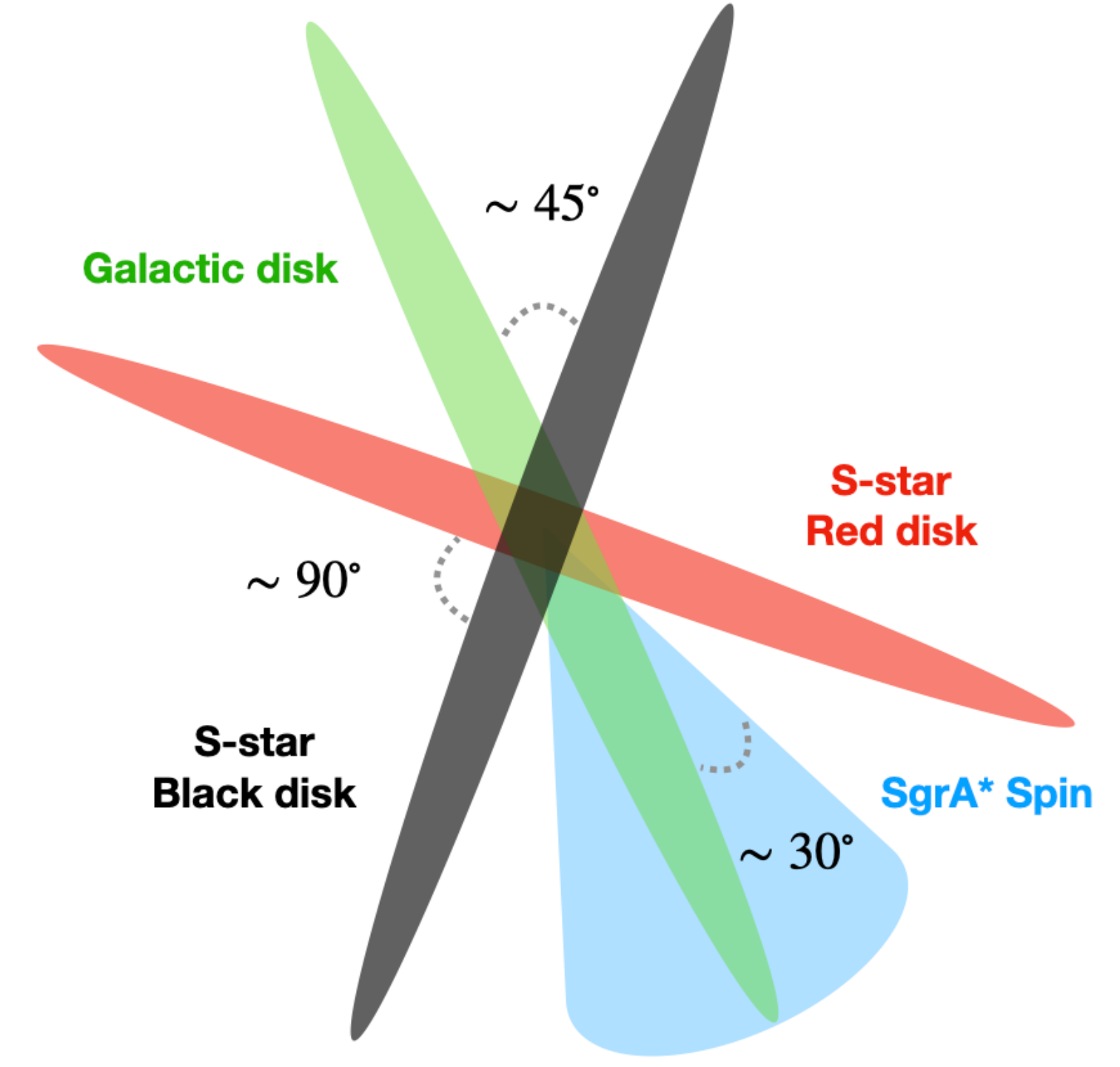}
\caption{Schematic representation of the spatial configuration of the S-stars, the Galactic disk, and the spin of SgrA$^*$. Red and black disks of S-stars are roughly perpendicular to each other and inclined of about $45^\circ$ with respect to the Galactic disk \citep{ali2020}. The blue cone with an opening angle of about $30^\circ$ represents the most likely orientation of the spin of SgrA$^*$ \citep{AkiyamaAlberdi2022b}.}
\label{fig:figure}
\end{figure}

Figure~\ref{fig:figure} illustrates the geometrical configuration of the red and black disks of S-stars, respectively, according to the analysis of \citet{ali2020}. The two disks are nearly perpendicular to each other and inclined of about $45^\circ$ with respect to the Galactic disk. The blue cone with an opening angle of about $30^\circ$ represents the most likely orientation of the spin of SgrA$^*$ according to the EHT analysis \citep[see Figure 4 in][]{AkiyamaAlberdi2022}.

\section{Implications for the spatial distribution of S-stars}
\label{sect:spin}

The spin angular momentum $\boldsymbol{\mathcal{S}}$ of a SMBH of mass $M_{\rm BH}$ can be expressed in terms of the dimensionless spin $\chi_{\rm BH}=|\boldsymbol{\mathcal{S}}|(GM_{\rm BH}^2/c)^{-1}$, where $G$ is the gravitational constant and $c$ is the speed of light. If the pericenter of the stars is close enough to the SMBH, its spin induces a Lense–Thirring (frame dragging) precession, which simultaneously affects the orbital inclination, the argument of periapsis, and the longitude of the ascending nodes \citep{lt1918}. These three Keplerian orbital elements change over a characteristic timescale \citep[e.g.,][]{merritt2013}
\begin{equation}
T_{\mathcal{S}}=\frac{1}{\nu_{\mathcal{S}}}=\frac{c^3 a^3(1-e^2)^{3/2}}{2\chi_{BH} G^2M_{\rm BH}^2}\,,
\label{eqn:ts}
\end{equation}
where $a$ and $e$ are the orbital semi-major axis and eccentricity of the star, respectively. According to Eq.~\ref{eqn:ts}, the inclination of any stellar orbit within the SMBH equatorial plane would not be affected by frame dragging, while its effect would be maximal for stars with orbits orthogonal to the SMBH spin.

\begin{figure} 
\centering
\includegraphics[scale=0.535]{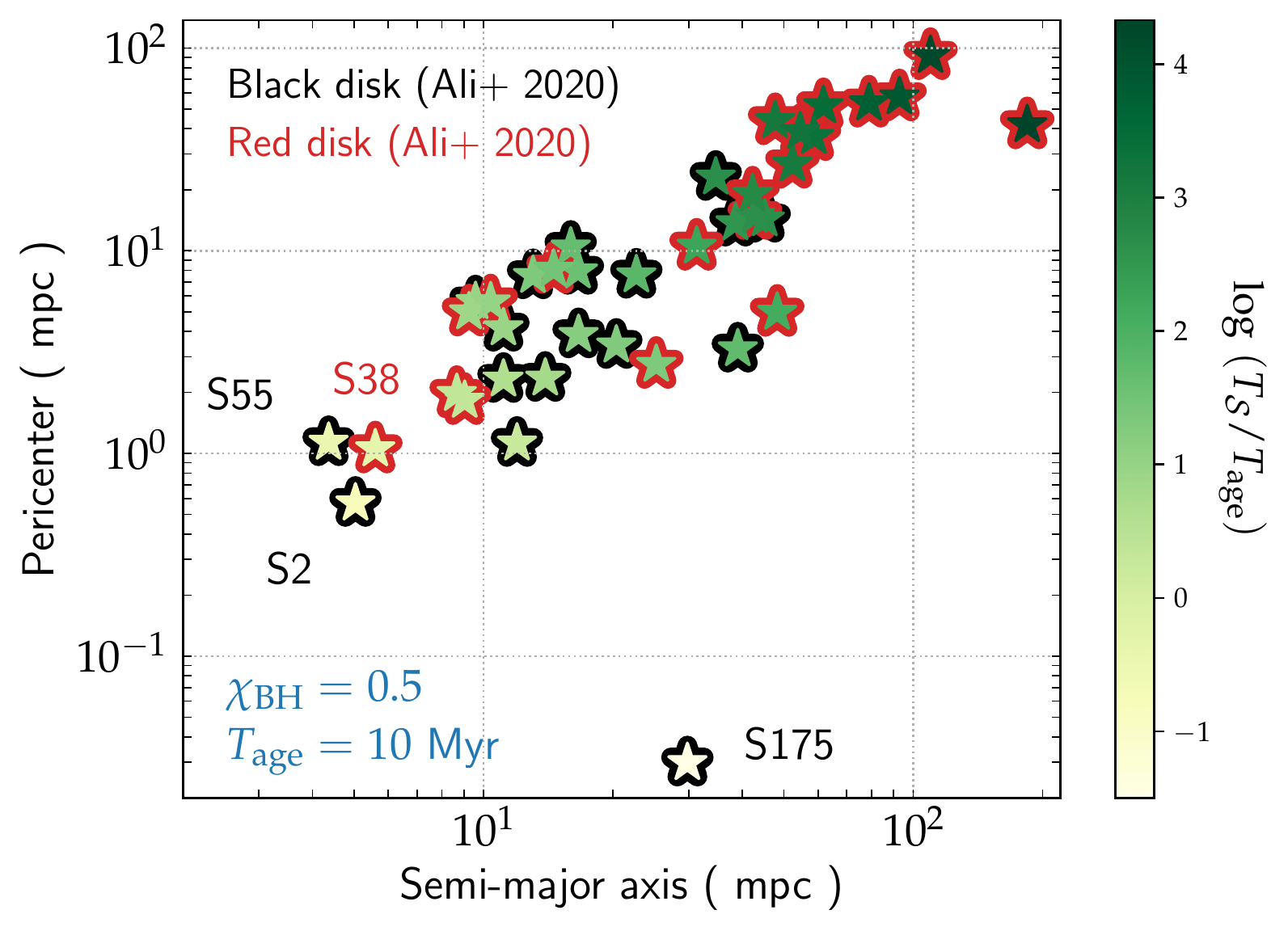}
\caption{Orbital distributions of S-stars according to \citet{ali2020}. Red and black symbols show the S-stars in the red and black disks, respectively. The color code represent the ratio between the frame-dragging timescale (Eq.~\ref{eqn:ts}), assuming the spin of SgrA$^*$ is $\chi_{\rm BH}=0.5$ \citep{AkiyamaAlberdi2022}, and the age of S-stars, assumed to be $10$~Myr \citep{habibi2017}.}
\label{fig:framdragg}
\end{figure}

The frame dragging precession by SgrA$^*$ could have a non-negligible effect on timescales that are shorter than main-sequence lifetimes of massive stars on stellar orbits within a milliparsec of the Galactic Center \citep{levin2003}. Only a handful of the known S-stars have pericenter passages from SgrA$^*$ small enough to be possibly affected by Lense-Thirring precession, assuming a non-zero spin of the SMBH. If the S-stars are arranged in two orthogonal disks as argued by \citet{ali2020}, the timescale for frame dragging precession of any star in the disks should be much longer than their age. Otherwise, the Lense-Thirring precession would have enough time to re-arrange the orbital inclinations of the S-stars, possibly erasing any disky signature. 

Using the previous argument and assuming that the S-stars have formed in the same plane in which we find them today, \citet{FragioneLoeb2020} showed that the spin of SgrA$^*$ can be constrained to be $\chi_{\rm BH}\lesssim 0.1$ if the classical S-stars (that is excluding the new six S-stars claimed by \citet{peiss2020a,peiss2020b}) are indeed organized in two orthogonal disks. However, the recent results by the EHT collaboration show a preference for $\chi_{\rm BH} >0.5$. Note that, since the SMBH is within about $30^\circ$ from the line of sight (see Figure~\ref{fig:figure}), this would imply that the frame dragging precession would be nearly maximal for one of the two disks of S-stars.

Figure~\ref{fig:framdragg} shows the timescale for frame dragging (Eq.~\ref{eqn:ts}) for the classical S-stars in the two orthogonal red and black disks according to the analysis of \citet{ali2020}. We assume that the spin of SgrA$^*$ is $\chi_{\rm BH}=0.5$ \citep{AkiyamaAlberdi2022} and that the age of S-stars is $10$~Myr \citep{habibi2017}. Four of the S-stars (S2, S38, S55, S175) have a Lense-Thirring precession timescale shorter than their lifetime. This implies that the frame dragging precession would have enough time to re-arrange the orbital inclinations of these four S-stars, driving them out of their current disks. Note that our argument holds whatever the relative inclination of the spin of SgrA$^*$ is with respect to the red and black disks. While S2, S55, and S175 belong to the black disk, S38 lies on the red disk. Therefore, their relative orbital inclination with respect to the SMBH spin could be either about $15^\circ$ or $75^\circ$, rendering the frame dragging precession nearly maximal for at least one of these four S-stars.

\begin{figure} 
\centering
\includegraphics[scale=0.575]{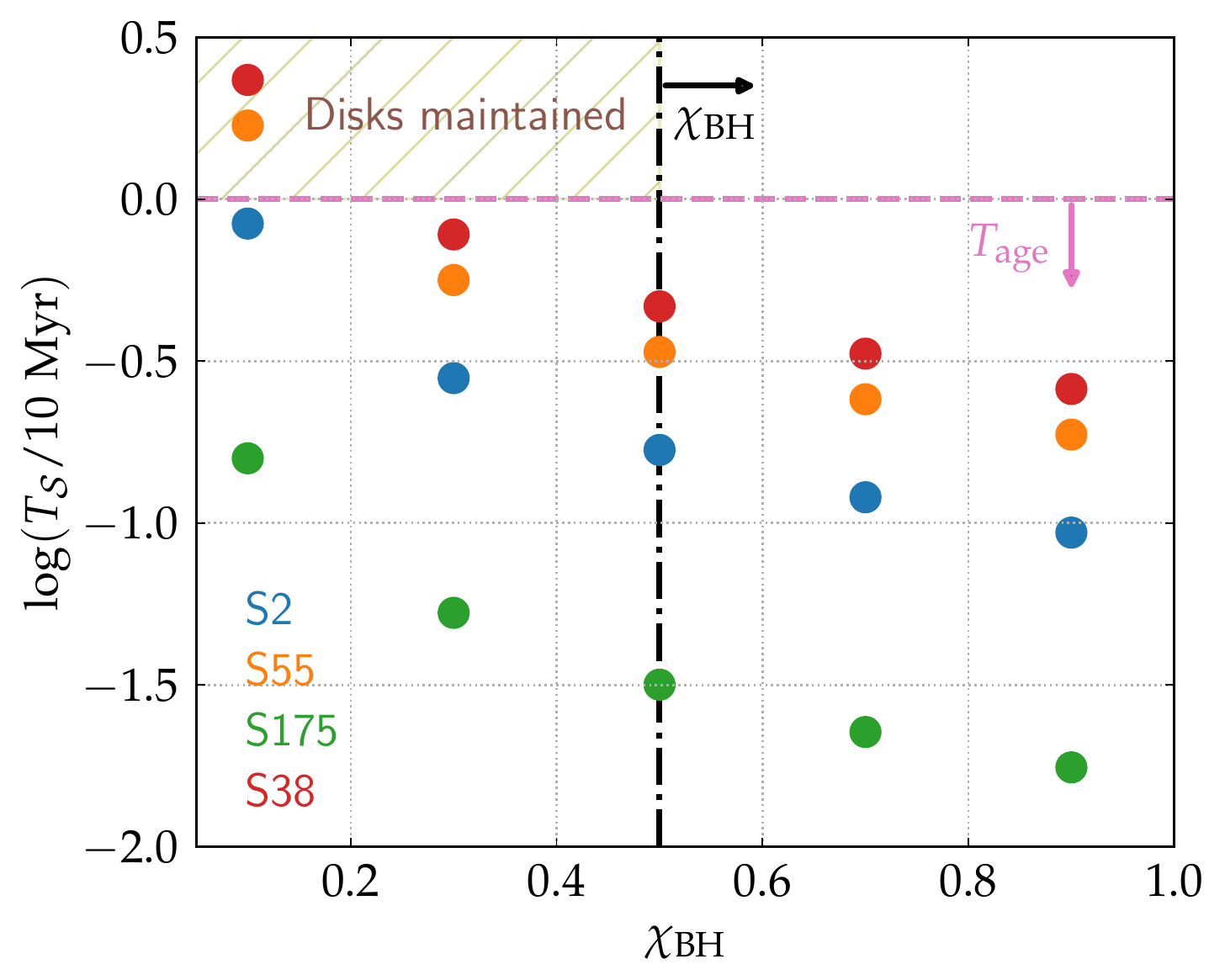}
\caption{S-stars (S2, S38, S55, S175) with frame-dragging timescale shorter than their age as a function of the spin of SgrA$^*$. The black line represents the minimum spin of SgrA$^*$ consistent with the EHT analysis \citep{AkiyamaAlberdi2022}, while the pink line represents the upper limit on the ages of S-stars from \citet{habibi2017}.}
\label{fig:framdragg2}
\end{figure}

To show how our results depend on the spin of SgrA$^*$ and the age of the S-stars, we show the timescale for frame dragging (Eq.~\ref{eqn:ts}) for S2, S38, S55, and S175 as a function of $\chi_{\rm BH}$. Only very small values of the SMBH spin ($\lesssim 0.1$) or very short stellar ages ($\lesssim 1$ Myr) would imply a precession timescale long enough to be consistent with these stars to be organized in two orthogonal disks. Note that including the newly claimed six S-stars by \citet{peiss2020a,peiss2020b} would make the discrepancy between the SMBH spin favored by the EHT analysis and the one required by the orbits of S-stars worse, owing to longer stellar ages of these fainter and less massive stars. 

\section{Conclusions}
\label{sect:concl}

Recently, the S-stars orbiting around SgrA$^*$ have been claimed to be arranged in two orthogonal disks \citep{ali2020}. In this case, the Lense-Thirring effect by the SMBH should not be strong enough to make their orbits precess and align them to the SMBH equatorial plane, thus requiring a negligible SMBH spin. However, the recent results by the EHT collaboration show a preference for a spin of SgrA$^*$ $\chi_{\rm BH} >0.5$.

We have reanalyzed the frame dragging precession for the classical S-stars and have showed that four of them (S2, S38, S55, and S175) have a Lense-Thirring timescale shorter than their lifetime. Only very small values of the SMBH spin ($\lesssim 0.1$) or very short stellar ages ($\lesssim 1$ Myr) would imply a precession timescale long enough to be consistent with these stars in two orthogonal disks.

Our results may have two possible implications. Assuming the estimate of the spin of SgrA$^*$ by the EHT collaboration holds, the S-stars could not be organized in two orthogonal disks. This would be consistent with the fact that the disky signature observed by \citet{ali2020} is stronger in projected coordinates rather than in physical ones, and with the fact that the S-stars seem to rotate in opposite directions in both disks, which is unlikely if the stars on the disks are coeval. Moreover, \citet{vonFellenbergGillessen2022} could not confirm that S-stars in the central region are organized in two orthogonal disks using GRAVITY data. Also, there could be older S-stars in a more isotropic distribution, which eluded detection owing to their lower luminosity. On the other hand, if the disky signature observed by \citet{ali2020} holds, the spin of SgrA$^*$ could not be too far from zero, in tension with the results of the EHT collaboration, which may suffer from uncertainties in theoretical modeling of the emission by the accreting gas and the resulting black hole silhouette.

\section*{Acknowledgements}

We thank Stefan Gillessen for useful discussions. G.F.\ acknowledges support from NASA Grant 80NSSC21K1722. This work was supported in part by Harvard's Black Hole Initiative, which is funded by grants from JFT and GBMF.

\bibliographystyle{yahapj}
\bibliography{refs}

\end{document}